\documentclass[12pt]{iopart}

\usepackage{graphicx,bm}

\newcommand{\be}{\begin{equation}}
\newcommand{\ee}{\end{equation}}
\newcommand{\bea}{\begin{eqnarray}}
\newcommand{\eea}{\end{eqnarray}}
\newcommand{\nn}{\nonumber \\}

\begin{document}

\tolerance=5000

\title{Modified gravity and its reconstruction from the universe expansion
history}

\author{Shin'ichi Nojiri\footnote{Electronic address:
nojiri@phys.nagoya-u.ac.jp}}
\address{Department of Physics, Nagoya University, Nagoya 464-8602. Japan}

\author{Sergei D. Odintsov\footnote{Electronic address: odintsov@ieec.uab.es,
also at TSPU, Tomsk}}
\address{Instituci\`{o} Catalana de Recerca i Estudis Avan\c{c}ats (ICREA)
and Institut de Ciencies de l'Espai (IEEC-CSIC),
Campus UAB, Facultat de Ciencies, Torre C5-Par-2a pl, E-08193 Bellaterra
(Barcelona), Spain}

\begin{abstract}

We develop the reconstruction program for the number of modified
gravities: scalar-tensor theory, $f(R)$, $F(G)$ and string-inspired,
scalar-Gauss-Bonnet gravity. The known (classical) universe expansion
history is used for the  explicit and successful reconstruction of some
versions (of special form or with specific potentials) from all above
modified gravities. It is demonstrated that
cosmological sequence of matter dominance, decceleration-acceleration
transition and acceleration era may always emerge as cosmological
solutions of such theory. Moreover, the late-time dark energy
FRW universe
may have the approximate or exact $\Lambda$CDM form consistent with
three years WMAP data. The principal possibility to extend this
reconstruction scheme to include  the radiation dominated era and
inflation is briefly mentioned. Finally, it is indicated how even modified
gravity which does not describe the matter-dominated epoch may have such a
solution before acceleration era at the price of the introduction of
compensating dark energy.

\end{abstract}

\maketitle

\section{Introduction}

The explanation of the current universe speed-up which is due to
mysterious dark
energy remains to be one of the
most fundamental  problems  of modern cosmology and theoretical physics.
The number of theoretical models have been developed aiming to describe
the dark energy universe (for recent review, see
\cite{reviewCST,padmanabhan}).
The obtained observational data indicate
that current universe has the effective equation of state parameter $w$
being very
close to $-1$ (for the review of observational data from the theoretical
 point of view, see \cite{bagla} and for description of observable
cosmological parameters, see \cite{tegmark}). Hence, it is still not
quite clear if
the universe lives in the phantom era ($w$ less than $-1$), in the (most
probably) cosmological constant epoch ($w=-1$) or in the quintessence
phase ($w$ more than $-1$). In such a situation, even phantom cosmological
models (see \cite{phantom,unification,tsujikawa} and refs. therein)
which show
quite strange properties are not ruled
out because even  being not phantomic one, the universe may currently
enter
to phantom dominated epoch.

The dark energy problem may have the gravitational origin, calling
for possible late-time modification of General Relativity. The number of
modified gravity models (for a review, see \cite{reviewNO}) starting from
the simplest $1/R$ theory \cite{c,CDTT} have been proposed as
gravitational alternative for dark energy. Such models may be inspired by
string/M-theory considerations \cite{no1,sasaki}. They may lead to
(phantom, cosmological constant or quintessence) effective cosmology at
late times without the need to introduce the scalar fields with strange
properties (like negative kinetic energy) or with complicated potentials.

Generally speaking, modified gravity looks very attractive as it gives
the qualitative answers to the number of fundamental questions about dark
energy.
Indeed, the origin of dark energy may be explained by some sub-leading
gravitational terms which become relevant with the decrease of
the curvature (at late times). Moreover, there are many proposals to
consider the gravitational terms relevant at high curvature (perhaps, due
to quantum gravity effects) as the source of the early-time inflation.
Hence, there appears the possibility to unify and to explain both: the
inflation and
late-time acceleration as the modified gravity  effects.
Meanwhile, at the intermediate epoch the gravity may be approximated by
General Relativity. Similarly, the coincidence problem may be explained.
It is remarkable that modified gravity may pretend to describe also dark
matter.

There are various ways to describe the modified gravity theory. Not only
$f(R)$  but also modified Gauss-Bonnet theory \cite{cognola} may be
presented in the form of scalar-tensor gravity (perhaps, with higher
derivative terms). Note that scalar-tensor theory itself may be understood
as some kind of modified gravity. Moreover, one can present modified
gravity as General
Relativity with quite complicated ideal fluid \cite{salvatore} having
non-standard equation of state. As it has been explained in
ref.\cite{mod},
this mathematical equivalence of different representations does not mean
the physical equivalence (due to transformed form of the physical metric
tensor). Nevertheless, it is remarkable that modified $f(R)$ gravity has
some close analogy with non-trivial $f(\rho)$ equation of state ideal
fluid \cite{EOS,tsujikawa}. In general, modified gravity may be considered
as General Relativity (GR) with inhomogeneous equation of state ideal
fluid \cite{impl2}. (The particular  example of this sort is given by ideal
fluid with time-dependent bulk viscosity \cite{brevik}).

Of course, modified gravity should be fitted against the observational
data (as is done in refs.\cite{constrain,hall}, mainly, for $f(R)$
theory).
Similarly,
it should pass the Solar System tests (for $f(R)$ models this is widely
discussed in refs.\cite{newton,dolgov,NO,hall} while SdS metrics are
studied in \cite{sds}). In principle, it may pass
the requested tests as some cancellation of the contributions between
different terms may occur as in the model of ref.\cite{NO}. Moreover, the
corresponding initial (boundary) conditions of the form of the action at
some (current) time of the universe  expansion history may be used to bring
the action to the approximate GR form (with possible earlier and (or) future
deviations from it).

In such a situation, any realistic
classical  gravitation
should describe the known sequence of the cosmological epochs (perhaps,
without the inflationary era where quantum gravity effects may be quite
strong). The purpose of this work is to review several versions of
modified gravity
and to show how it may be reconstructed from the known universe expansion
history (for recent discussion of reconstruction program for usual
gravity with
matter, see \cite{varun}). In the next section this problem is solved for
two scalar-tensor
theory. The reconstruction method is developed. Several examples of
realistic cosmology are considered: oscillating universe, $\Lambda$CDM
cosmology and asymptotically de Sitter space. The corresponding scalar
potentials admitting the sequence of matter dominated phase,
decceleration-acceleration transition and acceleration are constructed in
all above cases. Section three is devoted to the reconstruction of $f(R)$
gravity following ref.\cite{mod,IRGAC}. The  modified
gravity admitting the unification of matter dominated and acceleration
era is formulated. Its asymptotic behaviour at early and late times is
found. It is shown that it does not conflict with three years WMAP data.

In section four the alternative scenario for the model of ref.\cite{NO}
is developed. Adding the compensating dark energy which role is negligible
at the acceleration epoch it is shown that modified gravity \cite{NO}
may be viable (the matter dominated phase occurs before the cosmic
acceleration). In section five the reconstruction of string-inspired
scalar-Gauss-Bonnet gravity and of modified Gauss-Bonnet gravity is made.
Such theories have been suggested as dark energy models in
refs.\cite{sasaki,cognola}, respectively. Its versions where dark energy
universe emerges after matter dominance are given.
Some discussion and outlook are presented in the Discussion section.

\section{The reconstruction of scalar-tensor theory from the expansion history
of the universe}

\subsection{General formulation}

In this subsection, we show how {\it any} cosmology can be reproduced in the
scalar-tensor theory (or scalar-tensor theory may be reconstructed). Our
consideration is based on the method developed in ref.\cite{CNO} (for recent
study of scalar-tensor cosmology applied for dark energy description see
\cite{faraoni,tsujikawa} and for earlier attempts, see \cite{barrow}).
Note that studying the transition from non-phantom phase to phantom phase, the
instability of
the one scalar model becomes infinite at the transition point \cite{CNO}.
That is why we only consider two scalar-tensor theory, whose instability could
be always finite. Moreover, with two scalars it is easier to fit with universe
history expansion.

We now consider two scalar-tensor gravity
\be
\label{A1}
S=\int d^4 x \sqrt{-g}\left\{\frac{1}{2\kappa^2}R
  - \frac{1}{2}\omega(\phi)\partial_\mu \phi \partial^\mu \phi
  - \frac{1}{2}\eta(\chi)\partial_\mu \chi\partial^\mu \chi -
V(\phi,\chi)\right\}\ .
\ee
Here $\omega(\phi)$ is a function of the scalar field $\phi$ and $\eta(\chi)$
is a function
of the another scalar field  $\chi$.
It is assumed the spatially-flat FRW metric
$ds^2 = - dt^2 + a(t)^2 \sum_{i=1}^3 \left(dx^i\right)^2$ and
that $\phi$ and $\chi$ only depend on the time coordinate $t$.
The FRW equations lead to \be
\label{A2}
\omega(\phi) {\dot \phi}^2 + \eta(\chi) {\dot \chi}^2 = -
\frac{2}{\kappa^2}\dot H\ ,\quad
V(\phi,\chi)=\frac{1}{\kappa^2}\left(3H^2 + \dot H\right)\ .
\ee
Then if
\be
\label{A3}
\omega(t) + \eta(t)=- \frac{2}{\kappa^2}f'(t)\ ,\quad
V(t,t)=\frac{1}{\kappa^2}\left(3f(t)^2 + f'(t)\right)\ ,
\ee
the explicit solution follows
\be
\label{A4}
\phi=\chi=t\ ,\quad H=f(t)\ .
\ee
One may choose that $\omega$ should be always positive and $\eta$ be always
negative, for example
\bea
\label{A5}
\omega(\phi)&=&-\frac{2}{\kappa^2}\left\{f'(\phi) - \sqrt{\alpha^2 +
f'(\phi)^2} \right\}>0\ ,\nn
\eta(\chi)&=&-\frac{2}{\kappa^2}\sqrt{\alpha^2 + f'(\chi)^2}<0\ .
\eea
Here $\alpha$ is a constant. Define a new function $\tilde f(\phi,\chi)$ by
\be
\label{A6}
\tilde f(\phi,\chi)\equiv - \frac{\kappa^2}{2}\left(\int d\phi \omega(\phi)
+ \int d\chi \eta(\chi)\right)\ .
\ee
The constant of the integration could be fixed to require
\be
\label{A7}
\tilde f(t,t)=f(t)\ .
\ee
If $V(\phi,\chi)$ is given by using $\tilde f(\phi,\chi)$ as
\be
\label{A8}
V(\phi,\chi)=\frac{1}{\kappa^2}\left(3{\tilde f(\phi,\chi)}^2
+ \frac{\partial \tilde f(\phi,\chi)}{\partial \phi}
+ \frac{\partial \tilde f(\phi,\chi)}{\partial \chi} \right)\ ,
\ee
the FRW  and the scalar field equations are also satisfied:
\bea
\label{A9}
0&=&\omega(\phi)\ddot\phi + \frac{1}{2}\omega'(\phi) {\dot \phi}^2
+ 3H\omega(\phi)\dot\phi + \frac{\partial \tilde V(\phi,\chi)}{\partial \phi}\
, \nn
0&=&\eta(\chi)\ddot\chi + \frac{1}{2}\eta'(\chi) {\dot \chi}^2
+ 3H\eta(\chi)\dot\chi + \frac{\partial \tilde V(\phi,\chi)}{\partial \chi}\ .
\eea

We now investigate the (in)stability of the model.
By introducing the new quantities, $X_\phi$, $X_\chi$, and $Y$ as
\be
\label{A10}
X_\phi \equiv \dot \phi\ ,\quad X_\chi \equiv \dot \chi\ ,\quad
Y\equiv \frac{\tilde f(\phi,\chi)}{H} \ ,
\ee
the FRW equations and the scalar field equations (\ref{A9}) are:
\bea
\label{A11}
\frac{dX_\phi}{dN}&=& - \frac{\omega'(\phi)}{2H \omega(\phi)}\left(X_\phi^2 - 1
\right) - 3(X_\phi-Y)\ ,\nn
\frac{dX_\chi}{dN}&=& - \frac{\eta'(\chi)}{2H \eta(\chi)}\left(X_\chi^2 -
1\right)
  - 3(X_\chi-Y)\ ,\nn
\frac{dY}{dN}&=&\frac{1}{2\kappa^2H^2}\left\{X_\phi \left(X_\phi Y -1\right)
+ X_\chi\left(X_\chi Y -1\right)\right\}\ .
\eea
Here $d/dN\equiv H^{-1}d/dt$. In the solution (\ref{A4}),  $X_\phi=X_\chi=Y=1$.
The following perturbation may be considered
\be
\label{A12}
X_\phi=1+\delta X_\phi\ ,\quad X_\chi=1 + \delta X_\chi\ ,\quad Y=1 + \delta Y\
.
\ee
Hence
\bea
\label{A13}
&& \frac{d}{dN}\left(\begin{array}{c}
\delta X_\phi \\
\delta X_\chi \\
\delta Y
\end{array}\right)
= M \left(\begin{array}{c}
\delta X_\phi \\
\delta X_\chi \\
\delta Y
\end{array}\right)\ ,\nn
&& M \equiv \left(\begin{array}{ccc}
- \frac{\omega'(\phi)}{H\omega(\phi)} - 3 & 0 & 3 \\
0 & - \frac{\eta'(\chi)}{H\eta(\chi)} - 3 & 3 \\
\frac{1}{2\kappa^2H^2} & \frac{1}{2 \kappa^2 H^2} & \frac{1}{\kappa^2 H^2}
\end{array}\right)\ .
\eea
The eigenvalues of the matrix $M$ are given by solving the following eigenvalue
equation
\bea
\label{A14}
0&=& \left(\lambda + \frac{\omega'(\phi)}{H\omega(\phi)} + 3\right)
\left(\lambda + \frac{\eta'(\chi)}{H\eta(\chi)} + 3\right)
\left(\lambda - \frac{1}{\kappa^2 H^2}\right) \nn
&& + \frac{3}{2\kappa^2 H^2}\left(\lambda + \frac{\omega'(\phi)}{H\omega(\phi)}
+ 3\right)
+ \frac{3}{2\kappa^2 H^2}\left(\lambda + \frac{\eta'(\chi)}{H\eta(\chi)} +
3\right)\ .
\eea
The eigenvalues (\ref{A14}) for the two scalar model are clearly finite.
Hence, the instability could be finite.
In fact, right on the transition point where $\dot H=f'(t)=0$ and therefore
$f'(\phi)=f'(\chi)=0$, for the choice in (\ref{A5}), we find
\be
\label{AA1}
\omega(\phi)=-\eta(\chi)=\frac{2\alpha}{\kappa^2}\ ,\quad
\omega'(\phi)=-\frac{2\ddot H}{\kappa^2}\ ,\quad \eta'(\chi)=0\ .
\ee
Then the eigenvalue equation (\ref{A14}) reduces to
\bea
\label{AA2}
&& 0=\lambda^3 + \left(-A-B + 6\right) \lambda^2 + \left(AB - 3A - 3B +
9\right) \lambda
  - \frac{3}{2}AB + 9B\ ,\nn
&& A\equiv \frac{\ddot H}{\alpha}\ ,\quad B\equiv \frac{1}{\kappa^2 H^2}\ .
\eea
Here we have chosen $\alpha>0$. Then the eigenvalues are surely finite, which
shows that even if the solution (\ref{A4}) is not stable, the solution
has non-vanishing measure and therefore the transition from non-phantom phase
to phantom one can  occur. We should also note that the solution (\ref{A4})
can be in fact stable. For example, we consider the case
$A,B\to 0$.  Eq.(\ref{AA2}) further reduces to
\be
\label{AA3}
0=\lambda\left(\lambda + 3\right)^2\ .
\ee
Then the eigenvalues are given by $0$ and $-3$. Since there is no positive
eigenvalue, the solution
(\ref{A4}) is stable in the case.

It is not difficult to extend the above formulation to the multi-scalars model,
whose action
is given by
\be
\label{mA1}
S=\int d^4 x \sqrt{-g}\left\{\frac{1}{2\kappa^2}R
  - \frac{1}{2}\sum_i\omega_i(\phi_i)\partial_\mu \phi_i \partial^\mu \phi_i -
V(\phi_i)\right\}\ .
\ee
Here $\omega_i(\phi_i)$ is a function of the scalar field $\phi_i$.
We now choose $\omega_i$ to satisfy
\be
\label{mA2}
\sum_i(t)=-\frac{2}{\kappa^2}f'(t)
\ee
by a proper function. and we also choose the potential $V(\phi_i)$ by
\be
\label{mA3}
V(\phi_i)=\frac{1}{\kappa^2}\left(3\tilde f(\phi_i) +
\sum_i\frac{\partial\tilde f}{\partial \phi_i}\right)\ .
\ee
Here
\be
\label{mA4}
\tilde f(\phi_i)\equiv - \frac{\kappa^2}{2}\sum_i \int d\phi_i
\omega_i(\phi_i)\ .
\ee
The constant of the integration in (\ref{mA4}) is determined to satisfy
\be
\label{mA5}
\left. \tilde f(\phi_i) \right|_{{\rm all}\ \phi_i=t}=f(t)\ .
\ee
Then a solution of the FRW equations and the scalar field equations are given
by
\be
\label{mA6}
\phi_i=t\ ,\quad H=f(t)\ .
\ee
The rest consideration coincides with the one given for two-scalars model.

\subsection{The examples of scalar-tensor theory reconstruction}

We now consider some explicit examples in order to demonstrate
how the scalar-tensor theory can be reconstructed from known expansion history
of the universe using the above developed formulation.

As a first example, we consider the model given by
\be
\label{k11}
f(\phi)=h_0 + h_1 \sin (\nu \phi)\ .
\ee
Here it is assumed $h_0$, $h_1$, and $\nu$ are positive.
By choosing $\alpha = h_1\nu$ in (\ref{A5}), one finds
\bea
\label{Ae1}
&& \omega(\phi)=-\frac{2h_1\nu}{\kappa^2}\left\{\cos(\nu \phi) - \sqrt{1
+ \cos^2(\nu \phi)}\right\}\ ,\nn
&& \eta(\chi)=-\frac{2h_1}\nu{\kappa^2}\sqrt{1 + \cos^2(\nu \chi)}\ ,\nn
&& \tilde f(\phi,\chi) = h_0 + h_1 \sin(\nu\phi) - \frac{\sqrt{2}}{\nu}\left\{
E\left(1/\sqrt{2}, \nu \phi\right) - E\left(1/\sqrt{2}, \nu
\chi\right)\right\}\ .
\eea
Here $E(k,x)$ is the second kind elliptic integral defined by
\be
\label{Ae2}
E(k,x)=\int_0^x dx \sqrt{1 - k^2 \sin^2 x}\ .
\ee
Note that similar reconstruction in case of the single scalar-tensor theory was
presented in \cite{CNO}.
Eq.(\ref{k11}) shows that the Hubble rate $H$ is given by
\be
\label{k13}
H=h_0 + h_1 \sin (\nu t)\ ,
\ee
which is oscillating. When $h_0>h_1>0$, $H$ is always positive and the universe
is expanding.
Since
\be
\label{k14}
\dot H = h_1\nu \cos (\nu t)\ ,
\ee
when $h_1\nu>0$, $w_{\rm eff}$, which is defined by
\be
\label{FRW3k}
w_{\rm eff}=\frac{p}{\rho}= -1 - \frac{2\dot H}{3H^2}\ ,
\ee
is greater than $-1$ (non-phantom phase) when
\be
\label{k15}
\left(2n - \frac{1}{2}\right)\pi < \nu t < \left(2n + \frac{1}{2}\right)\pi \ ,
\ee
and less than $-1$ (phantom phase) when
\be
\label{k16}
\left(2n + \frac{1}{2}\right)\pi < \nu t < \left(2n + \frac{3}{2}\right)\pi \ .
\ee
In (\ref{k15}) and (\ref{k16}), $n$ is an integer.
Hence, in the model (\ref{k11}), there occur multiply oscillations
between phantom and non-phantom phases. It could be that our universe
currently corresponds to late-time acceleration phase in such
oscillatory regime (for recent reconstruction examples for  oscillatory
universe, see \cite{osc} and references therein).

The present universe is expanding with acceleration. On the other hand,
there occured the earlier matter-dominated period,
where the scale factor $a$ behaves as $a\sim t^{2/3}$.
Such behavior could be generated by dust in the
Einstein gravity. The baryons are dust and (cold) dark matter could be a dust.
The ratio of the baryons and the dark matter in the present universe could be
$1:5$ or $1:6$,
which should not be changed even in matter dominant
era. It is not clear what the dark matter is. For instance, the dark
matter might not be the real matter but some
(effective) artifact which appears in the modified/scalar-tensor gravity.
In the present section, it is assumed that not only dark energy
but also the dark matter may originate from the scalar field $\phi$.

We now investigate that the transition from the matter dominant
period to the acceleration period could be realized in the present
formulation (for single scalar-tensor theory, see \cite{NOS}). In the
following, the
contribution from matter could be neglected since the ratio of the matter with
the (effective) dark matter could be small.

First example is
\be
\label{STm12}
H=f(t)=g'(t)=g_0 + \frac{g_1}{t}\ .
\ee
When $t$ is large, the first term in (\ref{STm12}) dominates and
the Hubble rate $H$ becomes a constant.
Therefore, the universe is asymptotically deSitter space, which is an
accelerating universe (for recent examples of late-time accelerating cosmology
in scalar-tensor theory, see \cite{faraoni,tsujikawa}). On the
other hand, when $t$ is small, the second term in (\ref{STm12}) dominates
and the scale factor behaves as $a\sim t^{g_1}$. Therefore if $g_1=2/3$,
the matter-dominated period could be realized.
Here, one of scalars may be considered as usual matter.

Since
\be
\label{2s1}
f'(t)=-\frac{g_1}{t}<0\ ,
\ee
instead of (\ref{A5}), by using a positive constant $\alpha$, we may choose
\be
\label{2s2}
\omega(\phi)=-\frac{2\left(1+\alpha\right)}{\kappa^2}f'(\phi)>0\ ,\quad
\eta(\chi) = \frac{2\alpha}{\kappa^2}f'(\phi)<0\ ,
\ee
that is
\be
\label{2s3}
\omega(\phi)=\frac{2(1+\alpha)}{\kappa^2}\frac{g_1}{\phi^2}\ ,\quad
\eta(\chi)=\frac{2\alpha}{\kappa^2}\frac{g_1}{\phi^2}\ .
\ee
One should note that in the limit $\alpha\to 0$, the scalar field $\chi$
decouples and single scalar model follows.
Then one obtains
\bea
\label{2s4}
\tilde f(\phi,\chi) &=& g_0 + \frac{(1+\alpha)g_1}{\phi} - \frac{\alpha
g_1}{\chi}\ ,\nn
V(\phi.\chi) &=& \frac{1}{\kappa^2}\left\{3\left(g_0 +
\frac{(1+\alpha)g_1}{\phi} - \frac{\alpha g_1}{\chi}\right)^2
  - \frac{(1+\alpha)g_1}{\phi^2} + \frac{\alpha g_1}{\chi^2}\right\}\ .
\eea
Hence, for scalar-tensor theory with such potentials the matter dominated era
occurs before the acceleration epoch.

Before going to the second example,
we consider the Einstein gravity with cosmological constant and with
matter characterized by the EOS parameter $w$.
FRW equation has the following form:
\be
\label{LCDM1}
\frac{3}{\kappa^2}H^2 = \rho_0 a^{-3(1+w)} + \frac{3}{\kappa^2 l^2}\ .
\ee
Here $l^2$ is  the inverse cosmological constant.
The solution of (\ref{LCDM1}) is given by
\be
\label{LCDM2}
a=a_0\e^{g(t)}\ ,\quad
g(t)=\frac{2}{3(1+w)}\ln \left(\alpha \sinh \left(\frac{3(1+w)}{2l}\left(t -
t_0 \right)\right)\right)\ .
\ee
Here $t_0$ is a constant of the integration and
\be
\label{LCDM3}
\alpha^2\equiv \frac{1}{3}\kappa^2 l^2 \rho_0 a_0^{-3(1+w)}\ .
\ee
As the second example, we consider metric(\ref{LCDM2}) for
scalar-tensor theory.
Since
\bea
\label{2s5}
&& f(t) \equiv g'(t)=\frac{1}{l}\coth\left(\frac{3(1+w)}{2l}\left(t - t_0
\right)\right)\ ,\nn
&& f'(t)=g''(t)=- \frac{3(1+w)}{2l^2}\sinh^{-2}\left(\frac{3(1+w)}{2l}\left(t -
t_0 \right)\right)<0\ ,
\eea
it is convenient to use (\ref{2s2}) instead of (\ref{A5}).
Then one arrives at
\bea
\label{2s6}
\omega(\phi) &=& \frac{3(1+w)(1+\alpha)}{\kappa^2 l^2}
\sinh^{-2}\left(\frac{3(1+w)}{2l}\left(\phi - t_0 \right)\right) > 0\ ,\nn
\eta(\chi) &=& - \frac{3(1+w)\alpha}{\kappa^2 l^2}
\sinh^{-2}\left(\frac{3(1+w)}{2l}\left(\chi - t_0 \right)\right) < 0\ ,\nn
\tilde f(\phi,\chi) &=&
\frac{1+\alpha}{l}\coth\left(\frac{3(1+w)}{2l}\left(\phi - t_0 \right)\right)
  - \frac{\alpha}{l}\coth\left(\frac{3(1+w)}{2l}\left(\chi - t_0 \right)\right)
\ ,\nn
V(\phi,\chi) &=& \frac{1}{\kappa^2}\left[ \frac{3}{l^2}\left\{
(1+\alpha)\coth\left(\frac{3(1+w)}{2l}\left(\phi - t_0 \right)\right) \right.
\right. \nn
&& \left.  - \alpha \coth\left(\frac{3(1+w)}{2l}\left(\chi - t_0 \right)\right)
\right\}^2 \nn
&& - \frac{3(1+w)(1+\alpha)}{l^2} \sinh^{-2}\left(\frac{3(1+w)}{2l}\left(\phi -
t_0 \right)\right) \nn
&& \left. + \frac{3(1+w)\alpha}{l^2}\sinh^{-2}\left(\frac{3(1+w)}{2l}\left(\chi
- t_0 \right)\right) \right]\ .
\eea
Thus, in both examples, (\ref{STm12}) and (\ref{LCDM2}), there occurs
the matter dominated stage, the
transition from
matter dominated phase to acceleration phase and acceleration epoch. In
the acceleration phase,
in the above examples, the universe asymptotically
approaches to deSitter space. This does not conflict with WMAP data.
Indeed, three years WMAP data have been analyzed in Ref.\cite{Spergel}.
The combined analysis of WMAP with supernova Legacy
survey (SNLS) constrains the dark energy equation of state $w_{DE}$ pushing it
towards the cosmological constant. The marginalized best fit values of the
equation of state parameter at 68$\%$ confidance level
are given by $-1.14\leq w_{\rm DE} \leq -0.93$. In case of a prior that
universe is
flat, the combined data gives $-1.06 \leq w_{\rm DE} \leq -0.90 $.
In the examples (\ref{STm12}) and (\ref{LCDM2}),
the universe goes to asymptotically deSitter space, which gives
$w_{\rm DE}\to -1$, which does not, of course, conflict
with the above constraints.
Note, however, one needs to fine-tune $g_0$ in (\ref{STm12})
and $1/l$ in (\ref{LCDM2}) to be
$g_0\sim 1/l \sim 10^{-33}$ eV, in order to reproduce the observed Hubble rate
$H_0\sim 70$ km$\,$s$^{-1}$Mpc$^{-1}\sim 10^{-33}$ eV.

The final remark is in order. There is no problem to include into the
above reconstruction scenario usual matter (say, ideal fluid, dust, etc)
and (or) to find the scalar potentials corresponding to realistic
cosmology in multi-scalar case. Moreover, in the same way one may include
the radiation dominated epoch where quantum effects may be still neglected
in
the above scenario. In the next section it is shown how similar
reconstruction scheme may be developed for modified gravity.


\section{Reconstruction of modified gravity with the unification of
matter
dominated and accelerated phases}

\subsection{General formulation}

In the present section we develop the general
reconstruction scheme for modified gravity with $f(R)$ action.
It is shown how any cosmology may define the implicit form of the function
$f$. The starting action
of modified gravity is:
\be
\label{FR1}
S=\int d^4 x \left\{f(R) + {\cal L}_{\rm matter}\right\}\ .
\ee
First we consider the proper Hubble rate $H$, which describes
the evolution of the universe with
 matter dominance era and accelerating expansion (for discussion of
various accelerating cosmologies from above action, see
\cite{NO,cosmology,c,CDTT}).
It turns out that one can find $f(R)$-theory realizing such a
cosmology (with or without matter). The construction is not
explicit \cite{mod} and it is necessary to solve the second order
differential
equation and algebraic equation. It shows, however, that, at
least, in principle, we could obtain any cosmology by properly
reconstructing a function $f(R)$ on theoretical level.

The equivalent form of above action is
\be
\label{PQR1}
S=\int d^4 x \sqrt{-g} \left\{P(\phi) R + Q(\phi) + {\cal L}_{\rm
matter}\right\}\ .
\ee
Here $P$ and $Q$ are proper functions of the scalar field $\phi$
and ${\cal L}_{\rm matter}$ is the matter Lagrangian density.
Since the scalar field does not have a kinetic term, it may be regarded
  as an auxiliary field. In fact, by the variation over $\phi$,
it follows
\be
\label{PQR2}
0=P'(\phi)R + Q'(\phi)\ ,
\ee
which may be solved with respect to $\phi$:
\be
\label{PQR3}
\phi=\phi(R)\ .
\ee
By substituting (\ref{PQR3}) into (\ref{PQR1}), one obtains $f(R)$-gravity:
\be
\label{PQR4}
S=\int d^4 x \sqrt{-g} \left\{f(R) + {\cal L}_{\rm matter}\right\}\ , \quad
f(R) \equiv P(\phi(R)) R + Q(\phi(R))\ .
\ee
By the variation of the action (\ref{PQR1}) with respect to the
metric $g_{\mu\nu}$, we obtain
\be
\label{PQR5}
0=-\frac{1}{2}g_{\mu\nu}\left\{P(\phi) R + Q(\phi) \right\}
  - R_{\mu\nu} P(\phi) + \nabla_\mu \nabla_\nu P(\phi)
  - g_{\mu\nu} \nabla^2 P(\phi) + \frac{1}{2}T_{\mu\nu}\ .
\ee
The equations corresponding to standard spatially-flat FRW universe are
\bea
\label{PQR6}
0&=&-6 H^2 P(\phi) - Q(\phi) - 6H\frac{dP(\phi(t))}{dt} + \rho \ ,\\
\label{PQR7}
0&=&\left(4\dot H + 6H^2\right)P(\phi) + Q(\phi)
+ 2\frac{d^2 P(\phi(t))}{dt} + 4H\frac{d P(\phi(t))}{dt} + p\ .
\eea
By combining (\ref{PQR5}) and (\ref{PQR6}) and deleting $Q(\phi)$,
we find the following equation
\be
\label{PQR7b}
0=2\frac{d^2 P(\phi(t))}{dt^2} - 2 H \frac{dP(\phi(t))}{d\phi} + 4\dot H
P(\phi) + p + \rho\ .
\ee
As one can redefine the scalar field $\phi$ properly, we may choose
\be
\label{PQR8}
\phi=t\ .
\ee
It is assumed that $\rho$ and $p$ are the sum from the contribution of the
matters
with a constant equation of state parameters $w_i$.
Especially, when it is assumed a combination of the radiation and dust,
one gets the standard expression
\be
\label{PQR9}
\rho=\rho_{r0} a^{-4} + \rho_{d0} a^{-3}\ ,\quad p=\frac{\rho_{r0}}{3}a^{-4}\ ,
\ee
with constants $\rho_{r0}$ and $\rho_{d0}$. If the scale factor $a$
is given by a proper function $g(t)$ as
\be
\label{PQR10}
a=a_0\e^{g(t)}\ ,
\ee
with a constant $a_0$, Eq.(\ref{PQR7}) reduces
to the second rank differential equation:
\bea
\label{PQR11}
0 &=& 2 \frac{d^2 P(\phi)}{d\phi^2} - 2 g'(\phi) \frac{dP(\phi))}{d\phi} +
4g''(\phi) P(\phi) \nn
&& + \sum_i \left(1 + w_i\right) \rho_{i0} a_0^{-3(1+w_i)}
\e^{-3(1+w_i)g(\phi)} \ .
\eea
In principle, by solving (\ref{PQR11}) we find the form of $P(\phi)$.
Using (\ref{PQR6}) (or equivalently (\ref{PQR7})), we also find the form of
$Q(\phi)$ as
\be
\label{PQR12}
Q(\phi) = -6 \left(g'(\phi)\right)^2 P(\phi) - 6g'(\phi) \frac{dP(\phi)}{d\phi}
+ \sum_i \rho_{i0} a_0^{-3(1+w_i)}  \e^{-3(1+w_i)g(\phi)} \ .
\ee
Hence, in principle, any cosmology
expressed as (\ref{PQR10}) can be realized by some specific $f(R)$-gravity.

\subsection{ $f(R)$ gravity: the transition of matter dominated
phase  to the acceleration phase}

Let us consider  realistic example  where the total action contains
also usual matter. The starting form of  $g(\phi)$ is
\be
\label{PQR24}
g(\phi)=h(\phi) \ln \left(\frac{\phi}{\phi_0}\right)\ ,
\ee
with a constant $\phi_0$. It is assumed  that $h(\phi)$ is a slowly
changing function of $\phi$.
We use adiabatic approximation and neglect the derivatives
of $h(\phi)$ $\left(h'(\phi)\sim h''(\phi) \sim 0\right)$.
Then the solution of Eq.(\ref{PQR11}) has the following
form \cite{mod,IRGAC}:
\be
\label{PQR26}
P(\phi) = p_+ \phi^{n_+(\phi)} + p_- \phi^{n_-(\phi)} + \sum_i p_i(\phi)
\phi^{-3(1+w_i)h(\phi) + 2} \ .
\ee
Here $p_\pm$ are arbitrary constants and
\bea
\label{PQR27}
n_\pm (\phi) &\equiv& \frac{h(\phi) - 1 \pm \sqrt{h(\phi)^2 + 6h(\phi) +
1}}{2}\ ,\nn
p_i (\phi) &\equiv& - \left\{(1+w)\rho_{i0} a_0^{-3(1+w_i)}
\phi_0^{3(1+w)h(\phi)}\right\} \nn
&& \times \left\{6(1+w)(4+3w) h(\phi)^2 - 2 \left(13 + 9w\right)h(\phi) +
4\right\}^{-1}\ .
\eea
Especially for the radiation and dust, one has
\bea
\label{PP2}
p_r (\phi) & \equiv & - \frac{4\rho_{r0}\phi_0^{4h(\phi)} }{3a_0^4
\left( 40 h(\phi)^2 - 32 h(\phi) + 4\right)}\ ,\nn
p_d (\phi) & \equiv & - \frac{\rho_{d0}\phi_0^{3h(\phi)} }{a_0^3
\left( 24 h(\phi)^2 - 26 h(\phi) + 4\right)}\ .
\eea
We also find the form of $Q(\phi)$ as
\bea
\label{PQR28}
Q(\phi) &=& - 6h(\phi)p_+ \left(h(\phi) + n_+(\phi) \right) \phi^{n_+ (\phi) -
2} \nn
&& - 6h(\phi)p_- \left(h(\phi) + n_-(\phi) \right) \phi^{n_- (\phi) - 2} \nn
&& + \sum_i\left\{ - 6h(\phi) \left( -(2+3w)h(\phi) + 2\right)p_i (\phi)
\right. \nn
&& \left. + p_{i0} a_0^{-3(1+w)}\phi_0^{3(1+w)h(\phi)}\right\} \phi^{-
3(1+w)h(\phi)} \ .
\eea
Eq.(\ref{PQR24}) tells that $H\sim h(t)/t$ and
$R\sim 6\left( -h(t) + 2h(t)^2\right)/t^2$.
Let assume $\lim_{\phi\to 0} h(\phi) = h_i$ and $\lim_{\phi\to \infty} h(\phi)
= h_f$.
Then if $0<h_i<1$, the early universe is in decceleration phase and
if $h_f>1$, the late universe is in acceleration phase.
We may consider the case $h(\phi)\sim h_m$ is almost constant when $\phi\sim
t_m$
$\left(0\ll t_m \ll +\infty\right)$.
If $h_1$, $h_f>1$ and $0<h_m<1$,  the early universe is also accelerating,
which could correspond to the inflation. After that the universe
deccelerates,
which corresponds to matter-dominated phase with $h(\phi)\sim 2/3$ there.
Furthermore, after that, the universe could be in the acceleration phase.

The simplest example is
\be
\label{PQ1}
h(\phi) = \frac{h_i + h_f q \phi^2}{1 + q \phi^2}\ ,
\ee
with constants $h_i$, $h_f$, and $q$. When $\phi\to 0$, $h(\phi)\to h_i$ and
when $\phi\to \infty$, $h(\phi)\to h_f$. If $q$ is small enough, $h(\phi)$ can
be a slowly
varying function of $\phi$. Then we find \cite{mod,IRGAC}
\bea
\label{PQ2}
&& \phi^2=\Phi_0(R)\ ,\quad \Phi_0 \equiv \alpha_+^{1/3} + \alpha_-^{1/3}\ ,\nn
&& \alpha_\pm \equiv \frac{-\beta_0 \pm \sqrt{\beta_0^2 -
\frac{4\beta_1^3}{27}}}{2}\ ,\nn
&& \beta_0 \equiv \frac{2\left(2R + 6h_f q - 12 h_f^2 q \right)^3}{27 q^3 R^3}
\nn
&& \qquad - \frac{\left(2R + 6h_f q - 12 h_f^2 q \right)\left(R + 6h_i q + 6
h_f q - 4h_i h_f q\right)}{3qR} \nn
&& \qquad + 6h_i - 12 h_i^2\ ,\nn
&& \beta_1 \equiv - \frac{\left(2R + 6h_f q - 12 h_f^2 q \right)^2}{3 q^2 R^2}
  - \frac{R + 6h_i q + 6 h_f q - 4h_i h_f q}{q^2 R}\ .
\eea
There are two branches besides $\Phi_0$ but
the asymptotic behaviour of $R$ indicates that we should choose $\Phi_0$ in
(\ref{PQ2}). Then explicit form of $f(R)$ could be given by using
the expressions of $P(\phi)$ (\ref{PQR26}) and $Q(\phi)$  (\ref{PQR28}) as
\be
\label{PQ3}
f(R)=P\left(\sqrt{\Phi_0(R)}\right) R + Q\left(\sqrt{\Phi_0(R)}\right)\ .
\ee

One may check the asymptotic behavior of $f(R)$ in (\ref{PQ3}).
For simplicity, it is considered the case that the matter is only dust
($w=0$) and  that $p_-=0$. Then
\be
\label{nm1}
P(\phi) = p_+ \phi^{n_+(\phi)} + p_d(\phi) \phi^{-3h(\phi) + 2} \ .
\ee
One may always get $n_+ - \left(-3h + 2\right) >0$ in (\ref{nm1}).
Here $n_+$ is defined in (\ref{PQR27}).
Then when $\phi$ is large, the first term in (\ref{nm1}) dominates and
when $\phi$ is small, the last term dominates.
When $\phi$ is large, curvature is small and  $\phi^2\sim 6\left( - h_f + 2 h_f
\right)/R$
and $h(\phi)\to h(\infty)=h_f$.
Hence, Eq.(\ref{nm1}) shows that
\be
\label{PQasym1}
P(\phi) \sim p_+ \left(\frac{6\left( - h_f + 2 h_f
\right)}{R}\right)^{\left(h_f - 1
+ \sqrt{h_f^2 + 6h_f + 1}\right)/4} \ ,
\ee
and therefore $f(R)\sim R^{-\left(h(\phi) - 5 + \sqrt{h_f^2 + 6h_f +
1}\right)/4 }$.
Especially when $h\gg 1$, we find $f(R)\sim R^{-h_f/2}$.
Therefore there appears the negative power of $R$ (for review of such
theories, see \cite{reviewNO}). As $H\sim h_f/t$,
if $h_f>1$, the universe is in acceleration phase.

On the other hand, when curvature is large, we find
$\phi^2\sim 6\left( - h_i + 2 h_i \right)/R$ and $h(\phi)\to h(0)=h_i$.
Then (\ref{PQR26}) shows $P(\phi) \sim p_d(0) \phi^{-3h_i + 2}$.
If the universe era corresponds to matter dominated phase ($h_i=2/3$),
$P(\phi)$ becomes a constant and therefore $f(R)\sim R$, which reproduces the
Einstein gravity.

Thus, in the above model, matter dominated phase evolves into acceleration
phase and $f(R)$ behaves as $f(R)\sim R$ initially while
$f(R)\sim R^{-\left(h(\phi) - 5 + \sqrt{h_f^2 + 6h_f + 1}\right)/4 }$ at late
time. Moreover, for some parameter choice the asymptotic limit of above
theory reproduces the model of ref.\cite{NO}.

In our model, we can identify
\be
\label{w1}
w_{\rm DE} = -1 + \frac{2}{3 h_f}\ , \quad \mbox{or} \quad h_f=
\frac{2}{3\left(1 + w_{\rm DE}\right)}\ ,
\ee
which tells $h_f$ should be large if $h_f$ is positive. For example, if $w_{\rm
DE}=-0.93$, $h_f\sim 9.51\cdots$ and
if $w_{\rm DE}=-0.90$, $h_f\sim 6.67\cdots$.
Thus, we presented the example of $f(R)$ gravity which describes the
matter dominated stage, transition from decceleration to acceleration and
acceleration epoch which is consistent with three years WMAP.
Adding the radiation permits to realize the radiation dominated era before
above cosmological sequence as is shown in \cite{IRGAC}.

\section{Modified gravity and compensating dark energy}

In this section we  present another approach \cite{mod} to realistic
cosmology in
modified gravity.
Specifically, we discuss the modified gravity which successfully
describes the acceleration epoch but may be not viable in matter dominated
stage. In this case, it is demonstrated that one can introduce the
compensating dark energy (some ideal fluid) which helps to
realize matter dominated and decceleration-acceleration transition phases.
The role of such compensating dark energy is negligible in the
acceleration epoch.

We now start with general $f(R)$-gravity action (\ref{FR1}).
In the FRW metric with flat spatial dimensions one gets
\bea
\label{PQE53}
\rho &=& f(R) - 6\left(\dot H + H^2 - H \frac{d}{dt}\right)f'(R)\ ,\nn
p&=& - f(R) - 2 \left( - \dot H - 3 H^2 + \frac{d^2}{dt^2} + 2 H
\frac{d}{dt}\right) f'(R)\ .
\eea
Here $R= 6\dot H + 12 H^2$. If the Hubble rate is given (say, by observational
data)
as a function of $t$: $H=H(t)$, by substituting such expression into
(\ref{PQE53}),
we find the $t$-dependence of $\rho$ and $p$ as $\rho=\rho(t)$ and $p=p(t)$.
If one can solve the first equation with respect to $t$ as $t=t(\rho)$,
by substituting it into the second equation, an equation of state (EOS) follows
$p=p\left(t(\rho)\right)$. Of course, $\rho$ and $p$ could be a sum with the
contribution
of several kinds of fluids with simple EOS.

We now concentrate on the case that $f(R)$ is given by \cite{NO}
\be
\label{PQE55}
f(R)= - \frac{\alpha}{R^n} + \frac{R}{2\kappa^2} + \beta R^2\ .
\ee
Furthermore, we write $H(t)$ as $H(t)=h(t)/t$ and assume $h(t)$ is slowly
varying
function of $t$ and neglect the derivatives of $h(t)$ with respect to $t$.

First we consider the case that the last term in (\ref{PQE55}) dominates
$f(R)\sim \beta R^2$, which may correspond to the early (inflationary) epoch
of the universe. It is not  difficult to find
\be
\label{PQE57}
\rho \sim -\frac{36 \beta \left(-1 + 2h(t)\right) h(t)^2 }{t^{4n}}\ ,\quad
p \sim - \frac{36 \beta \left(-1 + 2h(t)\right) h \left(3h(t) +
1\right)}{t^{4n}}\ .
\ee
If $h$ goes to infinity, which corresponds to deSitter universe, we find
$\rho\sim p \sim h^3$ although $R\sim h^4$. Therefore $\rho$, $p  \ll \beta
R^2$ and
contribution form the matter could be neglected. Then the inflation could be
generated
only by the contribution from the higher curvature term.

Second, we consider the case that the second term in (\ref{PQE55})
dominates $f(R)\sim R/2\kappa^2$,
which may correspond to the matter dominated epoch after the inflation. In
this case $\rho$ and $p$ behave as
\be
\label{PQE58}
\rho \sim \frac{12 h(t) + 6h(t)^2}{\kappa^2 t^2}\ ,\quad
p \sim - \frac{4h(t) - 6h(t)^2}{\kappa^2 t^2}\ .
\ee
In the matter dominated epoch, we expect $h\sim 2/3$ $\left(a\sim
t^{\frac{2}{3}}\right)$. Hence, one gets $\rho \sim 32/3\kappa^2 t^2$, $p \sim
0$.
Therefore in the matter sector, dust with $w=0$ ($p=0$) should dominate, as
usually expected.

Finally we consider the case that the first term in (\ref{PQE55}) dominates
$f(R)\sim - \alpha/R^n$,
which might describe the acceleration of the present universe. The behavior
of $\rho$ and $p$
is given by
\bea
\label{PQE60}
\rho&\sim& \alpha\left\{6\left(n+1\right)\left(2n + 1\right)h(t) + 6
\left(n-2\right)h(t)^2\right\} \nn
&& \times \left\{-6h(t) + 12 h(t)^2 \right\}^{-n-1}t^{2n} \ ,\nn
p&\sim& \alpha\left\{ -4n\left(n+1\right)\left(2n+1\right) -2 \left(8 n^2 + 5n
+3\right)h(t)
   - 6 \left(n-2\right)h(t)^2\right\} \nn
&& \times \left\{-6h(t) + 12 h(t)^2 \right\}^{-n-1} t^{2n}\ .
\eea
Thus, the effective EOS parameter $w_l\equiv p/\rho$ is given by
\bea
\label{PQE61}
w_l &\sim & \left\{ -4n\left(n+1\right)\left(2n+1\right) -2 \left(8 n^2 + 5n
+3\right)h(t) \right. \nn
&& \left.  - 6 \left(n-2\right)h(t)^2\right\} \left\{6\left(n+1\right)\left(2n
+ 1\right)h(t)
+ 6 \left(n-2\right)h(t)^2\right\}^{-1}\ .
\eea
In order that the acceleration of the universe could occur, we find $h>1$.
Let us  now assume that $h(t)\to h_f$ when $t\to \infty$. Then one obtains $w_l
\to w_f$. Here
\bea
\label{PQE64}
w_f &\equiv& \left\{ -4n\left(n+1\right)\left(2n+1\right) -2 \left(8 n^2 + 5n
+3\right)h_f \right. \nn
&& \left.  - 6 \left(n-2\right)h_f^2\right\} \left\{6\left(n+1\right)\left(2n +
1\right)h_f
+ 6 \left(n-2\right)h_f^2\right\}^{-1}\ ,
\eea
and $H(t)\to h_f/t$. Since the matter energy density $\rho_{w_f}$ with the EoS
parameter $w_f$ behaves as
\be
\label{PQE65}
\rho_{w_f} \propto a^{-3(1+w_f)} \propto \exp \left( -3(1+w_f)\int dt H(t)
\right)\ ,
\ee
the energy density is
\be
\label{PQE66}
\rho_{w_f} \propto t^{-3(1+w_f)h_f}\ .
\ee
Comparing (\ref{PQE66}) with (\ref{PQE60}), we find $2n = -3(1+w_f)h_f$,
which can be confirmed directly from (\ref{PQE64}).

  From the above consideration, we find $\rho$ and $p$ contain mainly
contributions from dust with $w=0$,
$\rho_d(t)$, $p(t)=0$ and ``dark energy'' with $w=w_l$ in (\ref{PQE61}),
$\rho_l(t)$, $p_l(t)$. In the expressions
of $\rho(t)$ and $p(t)$ in (\ref{PQE53}), there might be a remaining part:
\be
\label{PQE62}
\rho_R(t) \equiv \rho(t) - \rho_d(t) - \rho_l(t)\ ,\quad
p_R(t) \equiv p(t) - p_l(t)\ ,
\ee
which may help the transition from the matter dominated epoch to the
acceleration epoch. By deleting $t$ in the
expression of (\ref{PQE62}), we obtain the EOS for the remaining part:
\be
\label{PQE63}
p_R=p_R(\rho_R)\ ,
\ee
which may be called the compensating dark energy.
More concretely, one may have
\be
\label{PQE68}
\rho_d \sim \frac{32}{3\kappa^2 t_0^2}\e^{-3\int_{t_0}^t dt \frac{h(t)}{t}}\ ,
\ee
and according to (\ref{PQE60}),
\bea
\label{PQE69}
\rho_l&\sim & \alpha\left\{6\left(n+1\right)\left(2n + 1\right)h_f + 6
\left(n-2\right)h_f^2\right\} \nn
&& \times \left\{-6h_f + 12 h_f^2
\right\}^{-n-1}t_1^{2n}\e^{-3(1+w_f)\int_{t_1}^t dt \frac{h(t)}{t}}\ .
\eea
In (\ref{PQE69}), we choose $t_1$ to be large enough. When $t\sim t_0$,
  $\rho(t) \sim \rho_d$
and when $t\to \infty$, $\rho(t) \sim \rho_l$. Thus, $\rho_R$ only
dominates after $t=t_1$ but it becomes
smaller in late times. Hence, the role of $\rho_R$ (which perhaps may be
identified partially with dark matter) is only to connect the matter
dominated epoch to the acceleration epoch.
The number of modified gravities can be made cosmologically viable via
such scenario.

\section{Reconstruction of scalar-Gauss-Bonnet theory and $F(G)$-gravity}

\subsection{Scalar-Gauss-Bonnet gravity}
In this section we
 show how string-inspired, scalar-Gauss-Bonnet gravity may be
reconstructed for any requested cosmology.  The starting action is
\be
\label{GBany1}
S=\int d^4 x \sqrt{-g}\left[ \frac{R}{2\kappa^2} - \frac{1}{2}\partial_\mu \phi
\partial^\mu \phi - V(\phi) - \xi_1(\phi) G \right]\ .
\ee
Here $G$ is the Gauss-Bonnet invariant and the scalar field $\phi$ is canonical
in (\ref{GBany1}). Note that scalar may be considered as matter component.
We assume the FRW universe and the scalar field $\phi$ only depending on $t$.
The FRW equations look like \cite{sasaki}:
\bea
\label{GBany4}
0&=& - \frac{3}{\kappa^2}H^2 + \frac{1}{2}{\dot\phi}^2 + V(\phi) + 24 H^3
\frac{d \xi_1(\phi(t))}{dt}\ ,\\
\label{GBany5}
0&=& \frac{1}{\kappa^2}\left(2\dot H + 3 H^2 \right) + \frac{1}{2}{\dot\phi}^2
- V(\phi)
  - 8H^2 \frac{d^2 \xi_1(\phi(t))}{dt^2} \nn
&& - 16H \dot H \frac{d\xi_1(\phi(t))}{dt} - 16 H^3 \frac{d
\xi_1(\phi(t))}{dt}\ .
\eea
and scalar field equation:
\be
\label{GBany6}
0=\ddot \phi + 3H\dot \phi + V'(\phi) + \xi_1'(\phi) G\ .
\ee
Now $G=24\left(\dot H H^2 + H^4\right)$.
Combining (\ref{GBany4}) and (\ref{GBany5}), one gets
\bea
\label{GBany7}
0&=&\frac{2}{\kappa^2}\dot H + {\dot\phi}^2 - 8H^2 \frac{d^2
\xi_1(\phi(t))}{dt^2}
  - 16 H\dot H \frac{d\xi_1(\phi(t))}{dt} + 8H^3 \frac{d\xi_1(\phi(t))}{dt} \nn
&=& \frac{2}{\kappa^2}\dot H + {\dot\phi}^2
  - 8a\frac{d}{dt}\left(\frac{H^2}{a}\frac{d\xi_1(\phi(t))}{dt}\right)\ .
\eea
Eq.(\ref{GBany7}) can be solved with respect to $\xi_1(\phi(t))$ as
\bea
\label{GBany8}
\xi_1(\phi(t)) &=& \frac{1}{8}\int^t dt_1 \frac{a(t_1)}{H(t_1)^2} W(t_1)\ ,\nn
W(t) &\equiv& \int^{t} \frac{dt_1}{a(t_1)}
\left(\frac{2}{\kappa^2}\dot H (t_1) + {\dot\phi(t_1)}^2 \right)\ .
\eea
Combining (\ref{GBany4}) and (\ref{GBany8}), the scalar potential $V(\phi(t))$
is:
\be
\label{GBany9}
V(\phi(t)) = \frac{3}{\kappa^2}H(t)^2 - \frac{1}{2}{\dot\phi (t)}^2  - 3a(t)
H(t) W(t)\ .
\ee
We now identify $t$ with $f(\phi)$ and $H$ with $g'(t)$ where $f$ and $g$
are some functions.
Let us consider the model where $V(\phi)$ and $\xi_1(\phi)$ may be
expressed
in terms of two functions $f$ and $g$ as
\bea
\label{GBany10b}
V(\phi) &=& \frac{3}{\kappa^2}g'\left(f(\phi)\right)^2 - \frac{1}{2f'(\phi)^2}
  - 3g'\left(f(\phi)\right) \e^{g\left(f(\phi)\right)} U(\phi) \, \nn
\xi_1(\phi) &=& \frac{1}{8}\int^\phi d\phi_1 \frac{f'(\phi_1)
\e^{g\left(f(\phi_1)\right)} }{g'\left(f(\phi_1)\right)^2} U(\phi_1)\ ,\nn
U(\phi) &\equiv& \int^\phi d\phi_1 f'(\phi_1 ) \e^{-g\left(f(\phi_1)\right)}
\left(\frac{2}{\kappa^2}g''\left(f(\phi_1)\right) + \frac{1}{f'(\phi_1 )^2}
\right)\ .
\eea
By choosing $V(\phi)$ and $\xi_1(\phi)$ as (\ref{GBany10b}), we can easily find
the following solution for Eqs.(\ref{GBany4}) and (\ref{GBany5}):
\be
\label{GBany11b}
\phi=f^{-1}(t)\quad \left(t=f(\phi)\right)\ ,\quad
a=a_0\e^{g(t)}\ \left(H= g'(t)\right)\ .
\ee
One can straightforwardly check the solution (\ref{GBany11b}) satisfies
the
field equation (\ref{GBany6}).

An interesting cosmological example is
\be
\label{LGB1}
\e^{g(t)}=\left(\frac{t}{t_0}\right)^{g_1}\e^{g_0 t}\ , \quad
\phi=f^{-1}(t)=\phi_0 \ln \frac{t}{t_0}\ .
\ee
Here $t_0$, $g_0$, $g_1$, and $\phi_0$ are constants. We further choose
\be
\label{LGB2}
\phi_0^2=\frac{2g_1}{\kappa^2}\ .
\ee
Then  $U$ is a constant: $U=U_0$ and
\bea
\label{LGB3}
V(\phi) &=& \frac{3}{\kappa^2}\left( g_0 + \frac{g_1}{t_0}\e^{-
\phi/\phi_0}\right)^2
  - \frac{g_1}{\kappa^2t_0^2} \e^{- 2\phi/\phi_0} \nn
&& - U_0\left(g_0 + \frac{g_1}{\phi}\right)\e^{- g_1 \phi/\phi_0}
\left(\frac{t}{t_0}\right)^{g_1} \e^{g_0 t_0\e^{\phi/\phi_0}}\ ,\nn
\xi_1 (\phi) &=& \frac{U_0}{8}\int^{t_0 \e^{\phi/\phi_0}} dt_1 \left( g_0 +
\frac{g_1}{t_1}\right)^{-2}
\left(\frac{t}{t_0}\right)\e^{g_0 t}\ .
\eea
Eq.(\ref{LGB2}) shows
\be
\label{LGB4}
H=g_0 + \frac{g_1}{t}\ .
\ee
Hence,, when $t$ is small, the second term in (\ref{LGB4}) dominates
and the scale factor behaves as $a\sim t^{g_1}$. Therefore if $g_1=2/3$,
the matter-dominated period could be realized.
On the other hand, when $t$ is large, the first term in (\ref{LGB4}) dominates
and
the Hubble rate $H$ becomes a constant.
Therefore, the universe is asymptotically deSitter space, which is an
accelerating universe.
As in our model the universe goes to asymptotically deSitter space, we find
$w_{DE} \to -1$.
Therefore our model can easily accommodate three years WMAP values of
$w_{DE}$.
For example, if $g_0 \simeq 40$, one has $w_{DE}=-0.98$.

In the limit $U_0\to 0$, the Gauss-Bonnet term in (\ref{GBany1}) vanishes and
the action (\ref{GBany1})
reduces into that of the usual scalar tensor theory with potential
\be
\label{LGB5}
V(\phi) = \frac{3}{\kappa^2}\left( g_0 + \frac{g_1}{t_0}\e^{-
\phi/\phi_0}\right)^2
  - \frac{g_1}{\kappa^2t_0^2} \e^{- 2\phi/\phi_0}\ ,
\ee
which reproduces the result \cite{NOS}.

Let us reconstruct  the scalar-Gauss-Bonnet gravity from FRW cosmology
(\ref{LCDM2}).
If a function $g(t)$ is given by (\ref{LCDM2}) and $f(\phi)$ is given by
\be
\label{LGB6}
f(\phi) = t_0 - \frac{2l}{3(1+w)}\ln\tanh \left( -
\frac{\kappa\sqrt{3(1+w)}}{4}\phi \right)\ ,
\ee
$U(\phi)$  (\ref{GBany10b}) becomes a constant again, $U=U(\phi)$. Then
$V(\phi)$ and $\xi(\phi)$ are
found to be
\bea
\label{LGB7}
V(\phi)&=& \frac{3}{\kappa^2 l^2}
\cosh^2\left(\frac{\kappa\sqrt{3(1+w)}}{2}\phi \right) \nn
&& - \frac{3(1+w)}{2l^2 \kappa^2}
\sinh^2\left(\frac{\kappa\sqrt{3(1+w)}}{2}\phi \right) \nn
&& - \frac{3U_0}{l}\cosh\left(\frac{\kappa\sqrt{3(1+w)}}{2}\phi \right) \nn
&& \times \left\{ - \frac{1}{\alpha}
\sinh\left(\frac{\kappa\sqrt{3(1+w)}}{2}\phi
\right)\right\}^{-2/\left(3(1+w)\right)}\ ,\nn
\xi_1(\phi)&=& - \frac{\alpha U_0 l^3 \kappa}{8\sqrt{3(1+w)}}\int^\phi d\phi_1
\cosh^{-2}\left(\frac{\kappa\sqrt{3(1+w)}}{2}\phi_1 \right) \nn
&& \times \left\{ - \frac{1}{\alpha}
\sinh\left(\frac{\kappa\sqrt{3(1+w)}}{2}\phi_1
\right)\right\}^{-2/\left(3(1+w)\right) - 1}\ ,
\eea
which again reproduces the result of ref.\cite{NOS} in the limit of
$U_0\to 0$.

Eq.(\ref{LCDM2}) shows that when $t\sim t_0$, the scale factor behaves as
$a\sim \left(t - t_0\right)^{2/(3(1+w))}$.
Therefore if $w=0$, the matter-dominated period could be realized.
On the other hand, when $t\to \infty$, $a$ behaves as $a\sim \e^{t/l}$, which
tells the universe goes to asymptotically
deSitter with $w_{\rm DE}\to -1$. Hence, it could be consistent with WMAP
data. The FRW cosmology of these and another versions of
scalar-Gauss-Bonnet gravity may be studied in the same way as it was done
in refs.\cite{sasaki1,sami}. The investigation of cosmological
perturbations can be also done \cite{koivisto} in the above model
admitting the cosmological sequence of matter dominance,
decceleration-acceleration transition and acceleration.

\subsection{$F(G)$-gravity}

The formulation of the previous section can be extended to so-called
$F(G)$-gravity \cite{cognola},
whose action is given by
\be
\label{fG1}
S=\int d^4 x \sqrt{-g}\left[ \frac{R}{2\kappa^2} + F(G) \right]
\ee
The above action could be rewritten by introducing the auxilliary scalar field
$\phi$ as
\be
\label{fG2}
S=\int d^4 x \sqrt{-g}\left[ \frac{R}{2\kappa^2} - V(\phi) - \xi_1(\phi) G
\right]\ .
\ee
By the variation over $\phi$, one obtains
\be
\label{fG3}
0=V'(\phi) + \xi_1'(\phi) G\ ,
\ee
which could be solved with respect to $\phi$ as
\be
\label{fG4}
\phi= \phi(G)\ .
\ee
By substituting the expression  (\ref{fG4}) into the action (\ref{fG2}), we
obtain the action of
$F(G)$-gravity with
\be
\label{fG5}
F(G)= - V\left(\phi(G)\right) + \xi_1\left(\phi(G)\right)G\ .
\ee
Note that the action (\ref{fG2}) could be obtained also by dropping the
kinetic term of $\phi$
in the action (\ref{GBany1}).

Assuming the spatially-flat  FRW universe and the scalar field $\phi$ only
depending on $t$,
 the FRW equations corresponding to (\ref{GBany4}) and (\ref{GBany5}) are:
\bea
\label{fG6}
0&=& - \frac{3}{\kappa^2}H^2 + V(\phi) + 24 H^3 \frac{d \xi_1(\phi(t))}{dt}\
,\\
\label{fG7}
0&=& \frac{1}{\kappa^2}\left(2\dot H + 3 H^2 \right) - V(\phi)
  - 8H^2 \frac{d^2 \xi_1(\phi(t))}{dt^2} \nn
&& - 16H \dot H \frac{d\xi_1(\phi(t))}{dt} - 16 H^3 \frac{d
\xi_1(\phi(t))}{dt}\ .
\eea
Combining the above equations, one gets
\bea
\label{fG8}
0 &=& \frac{2}{\kappa^2}\dot H - 8H^2 \frac{d^2
\xi_1(\phi(t))}{dt^2} - 16 H\dot H \frac{d\xi_1(\phi(t))}{dt} + 8H^3
\frac{d\xi_1(\phi(t))}{dt} \nn
&=& \frac{2}{\kappa^2}\dot H -
8a\frac{d}{dt}\left(\frac{H^2}{a}\frac{d\xi_1(\phi(t))}{dt}\right)\ ,
\eea
which can be solved with respect to $\xi_1(\phi(t))$ as
\be
\label{fG9}
\xi_1(\phi(t))=\frac{1}{8}\int^t dt_1 \frac{a(t_1)}{H(t_1)^2} W(t_1)\ ,\quad
W(t)\equiv \frac{2}{\kappa^2} \int^{t} \frac{dt_1}{a(t_1)} \dot H (t_1)\ .
\ee
Combining (\ref{fG6}) and (\ref{fG9}),  the following expression of
$V(\phi(t))$ may be found:
\be
\label{fG10}
V(\phi(t)) = \frac{3}{\kappa^2}H(t)^2 - 3a(t) H(t) W(t)\ .
\ee
Due to  a freedom of the redefinition of the scalar field $\phi$
 we may identify $t$ with $\phi$.
Then one considers the model where $V(\phi)$ and $\xi_1(\phi)$ can be
expressed
in terms of a single function $g$ as
\bea
\label{fG11}
&& V(\phi) = \frac{3}{\kappa^2}g'\left(\phi\right)^2 - 3g'\left(\phi\right)
\e^{g\left(\phi\right)} U(\phi) \, \quad
\xi_1(\phi) = \frac{1}{8}\int^\phi d\phi_1 \frac{\e^{g\left(\phi_1\right)}
}{g'(\phi_1)^2} U(\phi_1)\ ,\nn
&& U(\phi) \equiv \frac{2}{\kappa^2}\int^\phi d\phi_1
\e^{-g\left(\phi_1\right)} g''\left(\phi_1\right) \ .
\eea
By choosing $V(\phi)$ and $\xi_1(\phi)$ as (\ref{fG11}), we can easily find
the following solution for Eqs.(\ref{fG6}) and (\ref{fG7}):
\be
\label{fGB12}
a=a_0\e^{g(t)}\ \left(H= g'(t)\right)\ .
\ee
Then we can reconstruct $F(G)$-gravity in the way very similar \cite{sami,
preparation} to the
scalar-Gauss-Bonnet theory in the previous sub-section.

Although the above formulation is very similar to that in the
scalar-Gauss-Bonnet theory, there could be some difference.
In the scalar-Gauss-Bonnet gravity, since the scalar field has a kinetic term,
the scalar field could propagate and there
could be a possibility to generate extra force besides the Newton force. On the
other hand, $F(G)$-gravity has no kinetic
term for scalar field and extra force could not be generated.
In fact, one can consider the perturbation around the deSitter background,
writing the metric as
$g_{\mu\nu}=g_{(0)\mu\nu} + h_{\mu\nu}$.
Here the Riemann tensor in the deSitter background is given by
\be
\label{GB35}
R_{(0)\mu\nu\rho\sigma}=H_0^2\left(g_{(0)\mu\rho}g_{(0)\nu\sigma}
  - g_{(0)\mu\sigma}g_{(0)\nu\rho}\right)\ .
\ee
The flat background corresponds to the limit of $H_0\to 0$.
For simplicity, if we choose the gauge conditions $g_{(0)}^{\mu\nu}
h_{\mu\nu}=\nabla_{(0)}^\mu h_{\mu\nu}=0$,
we find from the equation of motion without energy-momentum tensor,
\be
\label{GB38}
0=\frac{1}{4\kappa^2} \left( \nabla^2 h_{\mu\nu} - 2H_0^2 h_{\mu\nu}\right)\ .
\ee
Since the contribution form the Gauss-Bonnet term does not appear except the
length parameter $1/H_0$ of the deSitter space, the only propagating mode
should
be graviton in the $F(G)$-gravity. This also shows that Newton law is
effectively the same as in GR.

Finally, let us note that it is not difficult to extend the above
reconstruction scheme for modified Gauss-Bonnet gravity so that the
(ideal fluid) matter may be naturally included into the
formulation \cite{preparation}.

\section{Discussion}

In summary, the reconstruction program is developed for the number of
modified gravities: scalar-tensor theory, $f(R)$, $F(G)$ and
scalar-Gauss-Bonnet gravity. It is explicitly demonstrated which versions
of above theories may be reconstructed from the known universe expansion
history. Specifically, it is shown that the cosmological sequence of
matter dominance, decceleration-acceleration transition and acceleration
era may always emerge as the cosmological solutions of such modified
gravities. Moreover, it is explained that the (exact or approximated)
$\Lambda$CDM cosmology may be also the  solution of such
gravity theories. Several examples of corresponding reconstruction for it
as well
as for oscillating universe where $\Lambda$CDM dark energy describes the
one of the oscillation periods are given.
In principle, it is not difficult to include into this reconstruction
scheme also radiation dominated era and, perhaps, the inflationary epoch.
This will be studied elsewhere.

It is interesting that even if specific modified gravity which is well
suited with acceleration epoch does not describe the matter dominance
period, such a period may be included there as the solution at the price of
the introduction of compensating dark energy.
The corresponding example is worked out for  $f(R)$ model of
ref.\cite{NO}. It may be also extended for scalar-Gauss-Bonnet or $F(G)$
gravity.

It is quite remarkable that modified gravity may not only naturally
describe dark
energy (unlike to conventional GR) but also it can be reconstructed using
the realistic  universe
expansion history. It is also promising that it may be fitted with
observational data. As well it may be compatible with Solar System tests.
Of course, some controversial results exist here what is not strange
due to fact that modified gravity is seriously considered as the
alternative cosmological theory only last several years while GR played
this role almost one century. One should also bear in mind that most of
cosmological data were obtained (at least, up to some extent) using GR
foundation. Hence, very serious reconsideration (as well as more precise
and
complete
observational data) are requested in order to have the answer to the
fundamental question: what is the gravitation theory which governs  the
expansion of our universe?

\section*{Acknowledgements}

We are grateful to G. Allemandi, I. Brevik, A. Borowiec, S. Capozziello,
G. Cognola, E.
Elizalde, M. Francaviglia, P. Gonzalez-Diaz, V.V. Obukhov, K.E. Osetrin,
S. Tsujikawa, H.
Stefancic, M. Sami and S. Zerbini for useful discussions  and (or)
collaboration on related questions.
We thank the organizers of Spanish ERE2006 conference
for the invitation to give a talk there. The work by S.N. was supported in
part by Monbusho grant no.18549001 (Japan) and XXI century COE program of
Nagoya university provided by JSPS (15COEEG01) and that by S.D.O. by the
project FIS2005-01181 (MEC, Spain)
 and by RFBR grant 06-01-00609 (Russia).

\end{document}